\documentclass[12pt]{article}
\usepackage{epsfig}
\usepackage{amsfonts}
\usepackage{amsopn}
\usepackage{amsmath}

\title{On the Equivalence Principle and Electrodynamics of Moving Bodies}

\author{Maciej Trzetrzelewski \thanks{e-mail: maciej.trzetrzelewski@gmail.com} \\ \\
M. Smoluchowski Institute of Physics, \\
Jagiellonian University, \\
\L{}ojasiewicza, St. 11, 30-348 Krak\'ow, \\
Poland\\ \\
NORDITA,  \\
Roslagstullsbacken 23, 106 91 Stockholm, \\
Sweden\\ \\
 Department of Mathematics,\\
Royal Institute of Technology, \\
KTH, 100 44 Stockholm, \\
Sweden
}

\begin{document}
\date{}
\maketitle

\abstract{
We consider a certain extension of the  Einstein's elevator thought experiment by assuming that the elevator is charged and falls into an electromagnetic field.
We argue, on grounds of the Equivalence Principle, that an observer-dependent metric should exist, for which the geodesics coincide with the trajectories of the charged body in the electromagnetic field. 
We give a solution to this problem by finding such a metric and hence we are able to recover the Lorentz-force law. Recovery of the leading terms of the radiation reaction force is also possible with this approach.
}

\section{Introduction}

In this paper we consider a particular modification of Einstein's elevator thought experiment. We will assume that the elevator is charged and falling freely with an observer inside it. It follows that there is no electromagnetic field inside the elevator; moreover, the observer cannot even tell if the elevator is indeed charged. At this stage the relative distance between the observer and the elevators walls are constant in time.

We then assume that the elevator falls into a region with an electric field and so starts accelerating towards the observer; i.e. the relative distance between the observer and one on the elevator's walls starts decreasing. At some point one of the elevator's walls will hit the observer and the observer will start accelerating with the same trajectory as the elevator. Entering the electromagnetic field is the real cause of this acceleration. However, the observer cannot conclude this unambiguously as he cannot detect the electromagnetic field locally. The observer may suspect that the elevator is charged and that it has entered a region with an electric field, but he may as well conclude that the elevator has hit another body (e.g. it has landed), which would explain why the observer now accelerates towards one side of the elevator.  We have therefore arrived at the following consequence of the equivalence principle: under the above assumptions, the observer cannot locally distinguish between a (fictitious) gravitational filed and a (real) electromagnetic one.

\section{Equivalence of equations of motion}

From the charged elevator thought experiment, it follows that the trajectory given by the Lorentz force equation should coincide with the geodesic in an appropriately chosen metric depending on the electromagnetic potential.
Therefore, we will be looking for a metric $g_{\mu\nu}$ which depends on the electromagnetic field $A_{\mu}$  in such a way that the equations for the geodesics imply equations of motion for a charged particle in the electromagnetic field. In doing so we are allowed to use the dimensional parameters of the problem i.e. the charge $q$ and the mass $m$ of the body. The anticipated metric has to depend on $q$ and $m$ if we want to derive the Lorentz force equation from the geodesic one, therefore the metric we are looking for will be observer dependent. This is not a concern since the metric $g_{\mu\nu}$ is fictitious and, by construction (of the thought experiment), may depend on the observer's parameters.

 Because the metric should be symmetric in indices a natural guess is
\begin{equation}  \label{metric}
g_{\mu\nu} = \eta_{\mu\nu} +  k\frac{q^2}{m^2} A_{\mu} A_{\nu},
\end{equation}
where the coefficient $q^2/m^2$ is chosen such that the metric is dimensionless (we work with $c=1$ units), $k$ is some dimensionless parameter, and the signature of the metric is $(+,-,-,-)$. The metric (\ref{metric}) is observer dependent i.e. it depends on the ratio $q/m$. Therefore, observers with different $q/m$ ratios will not agree on the metric. This however does not raise any issues since the thought experiment with the charged elevator does not imply that there should be unique metric for all observers.

A reader familiar with the Kaluza-Klein ansatz \cite{KK} will notice that (\ref{metric}) is almost an identical one, therefore  one might suspect that we are making a connection to the Kaluza-Klein theory. However, there are differences: most importantly, our approach is very much rooted in the Equivalence Principle in 4D. We are not attempting to geometrize electromdynamics, but rather to present it from a different view in 4D. Clearly, unlike in Kaluza-Klein theory, we are not making any unverified assumptions about the dimensionality of space-time or new field content/dynamics 
 (in the Kaluza-Klein approach, a counterpart of $k\frac{q^2}{m^2}$ in (\ref{metric}) is $g_{55}$ - the 5th diagonal component of the 5D metric). One might say that the price we are paying for this 4D picture is the universality of the metric i.e. the metric (\ref{metric}) is observer dependent. There are however additional benefits of our approach as we are able to recover not only the Lorentz-force law (as will be shown in this section) but also the leading terms of the radiation reaction force (see Section 7).

Note also that the metric (\ref{metric}) is not gauge invariant. This signals that the ansatz (\ref{metric}) is already in a certain gauge, i.e. the gauge for $A_{\mu}$ on the trajectory has already been chosen. In this chapter we will show that the equivalence of geodesic and Lorentz force equations requires a certain condition on $A_{\mu}$ on the trajectory. In the next chapter we will show that this condition is achievable by gauge transformation and hence confirm that the form of the metric as in (\ref{metric}) indeed fixes, albeit implicitly, the gauge choice of $A_{\mu}$ in this problem.

We will now derive a condition under which the equations of motion for the action
\begin{equation} \label{action}
S = - m \int \sqrt{g_{\mu\nu}u^{\mu} u^{\nu}}d\tau,
\end{equation}
where $u^{\mu} = \dot{x}^{\mu}$ is the four-velocity, are equivalent to the Lorentz force equation. Performing the variation of $S$ w.r.t. $x_{\mu}$ we find
\begin{equation} \label{kappa1}
\left(\frac{\eta_{\mu\nu}u^{\nu}}{\sqrt{gu^2}} \right)^.=  k \frac{\kappa}{\sqrt{gu^2}}\frac{q}{m}  F_{\mu\nu}u^{\nu} - k\frac{q}{m}\left(\frac{\kappa}{\sqrt{gu^2}}\right)^.A_{\mu},
\end{equation}
$$
gu^2:=g_{\mu\nu}u^{\mu} u^{\nu}, \ \ \ \ \kappa := \frac{q}{m}A_{\mu} u^{\mu},
$$
where we introduced the symbol $gu^2$ and a key dimensionless quantity $\kappa$. This result should be compared with the corresponding action for a 
charged particle and the resulting equations of motion
\begin{equation} \label{action1}
S_{A} = - m \int \sqrt{\eta_{\mu\nu}u^{\mu} u^{\nu}}d\tau   - q \int A_{\mu} u^{\mu}d\tau,
\end{equation}
\begin{equation} \label{eom}
\delta_x S_A=0 \ \ \ \Longrightarrow  \ \ \ \ \left(\frac{\eta_{\mu\nu}u^{\nu}}{\sqrt{u^2}} \right)^.=  \frac{q}{m}  F_{\mu\nu}u^{\nu}.
\end{equation}

Note that, since we are in curved space-time, the  lowering or raising of the induces should be done by the metric $g_{\mu\nu}$ or $g^{\mu\nu}$, which is why we kept  $\eta_{\mu\nu}$ explicitly in (\ref{kappa1}). On the other hand, we want the resulting equations of motion to be equivalent to Lorentz force equations written in Minkowski space. In practice, expressions like $g^{\mu\nu}A_{\mu}$ will appear rarely, while ones like $\eta^{\mu\nu}A_{\mu}$ will appear quite often. Therefore, in this paper we will adopt a non-standard convention and use $\eta_{\mu\nu}$/ $\eta^{\mu\nu}$ to lower/raise indices of $A_{\mu}, p_{\mu}, x^{\mu}$ and $F_{\mu\nu}$. The indices of $g_{\mu\nu}$ will never be raised or lowered; - $g_{\mu\nu}$ will always be written explicitly while $g^{\mu\nu}$ is defined as a reciprocal of $g_{\mu\nu}$.

  In order to have (\ref{kappa1}) equivalent to the Lorentz force equation we should at least get rid of the gauge-dependent term on the RHS of the equation (\ref{kappa1}). Therefore, we set
\begin{equation} \label{condition}
 \frac{\kappa}{\sqrt{gu^2}} = C,
\end{equation}
where $C$ is a constant. It is tempting to set $C=1/k$ so that the RHS of (\ref{kappa1}) is exactly equal to the Lorentz force, but in fact we should not do so because on the LHS we still have the incorrect factor $\sqrt{gu^2}$ instead of $\sqrt{u^2}$. The definition of $gu^2$ and $\kappa$ imply that $gu^2 = u^2 + k \kappa^2$; therefore the condition (\ref{condition}) gives  $ \kappa^2 = C^2 u^2/(1-kC^2) $,  which substituted to (\ref{condition}) results in

\begin{equation} \label{proportional}
\sqrt{gu^2} = \frac{1}{\sqrt{1-kC^2}}\sqrt{u^2}.
\end{equation}
The equations of motion are now
\begin{equation}  \label{kappa2}
\dot{p}_{\mu} = \frac{kC}{\sqrt{1-kC^2}} q F_{\mu\nu}u^{\nu}, \ \ \ \ p_{\mu} = m u_{\mu}/\sqrt{u^2},
\end{equation}
where we introduced momentum $p_{\mu}$. Now, to recover the Lorentz force law exactly, all we need to do is set $kC = \sqrt{1-kC^2}$, in which case the condition (\ref{condition}) becomes
\begin{equation}  \label{cons1}
q A_{\mu} p^{\mu} = m^2 \frac{1}{k}.
\end{equation}

Therefore, we have arrived at the main result of this paper. We have shown that there exists an $A_{\mu}$ dependent metric such that the corresponding geodesic equations are equivalent to the Lorentz force law equations. The existence of such a metric is a consequence of the thought experiment with the charged elevator. 
In the remaining part of the paper we will clarify the details regarding the gauge choice of $A_{\mu}$ as well as re-derive the above result directly from the equations of motions (not the action principle). We shall also investigate what the Einstein-Hilbert action reduces to, if metric (\ref{metric}) is used.
Note that if we now replace $\eta_{\mu\nu}$ in (\ref{metric}) with an arbitrary metric $h_{\mu\nu}$, we will arrive at the Lorentz force law in curved space corresponding to $h_{\mu\nu}$.

\section{A gauge choice}

Equations $ \kappa^2 = C^2 u^2/(1-kC^2) $ and (\ref{cons1}) imply that $C^2= 1/(k+k^2)$ and so
$ k \in (-\infty,-1)\cup (0,\infty)$ (because $C^2$ is positive and finite). However, the case $k=-1$ is interesting since (\ref{cons1}) can also be written as

\begin{equation} \label{lagran}
m\sqrt{u^2}- q k A_{\mu}u^{\mu} =0
\end{equation}
and so, for $k=-1$, (\ref{lagran}) is equivalent to saying that the Lagrangian of the charged particle vanishes on real trajectories. As we now show, this value is also distinguished from the point of view of gauge transformations.

The constraint (\ref{cons1}) is a necessary condition for the potential $A_{\mu}$ and needs to be satisfied on the trajectory of the particle, if the geodesics for the metric (\ref{metric}) are supposed to coincide with the trajectories of the charged particle in the electromagnetic field in the Minkowski space. This constraint is also a necessary one in order to make the equation (\ref{kappa1}) gauge invariant. Therefore, we can say that in order to maintain the gauge invariance of the equation (\ref{kappa1}), one nevertheless needs to fix the field $A_{\mu}$ in a marginal way, i.e. on the world-line of the particle. From this point of view, gauge invariance and equivalence principle are not independent concepts.

One can look at (\ref{cons1}) as some sort of gauge fixing; if $A'_{\mu}$ is an arbitrary potential then we can always make a gauge transformation
\begin{equation}  \label{gtr}
A_{\mu} = A'_{\mu} + \partial_{\mu}\chi
\end{equation}
 in such a way that (\ref{cons1}) will be satisfied. This is obtained by choosing a $\chi$ such that
$$
qA'_{\mu}u^{\mu} + q\dot{\chi} = m\sqrt{u^2}/k
$$
and so
\begin{equation}  \label{chi}
\chi  = \frac{1}{q } \int_{\gamma} \left(\frac{m}{k}\sqrt{u^2} - q A'_{\mu}u^{\mu}\right)d\tau,
\end{equation}
where $\gamma$ is the world-line of the particle. Therefore, for $k=-1$ we find that the phase $q\chi$ is in fact given by the action of the charged particle for the $A'_{\mu}$ field, evaluated at the classical trajectory i.e. by the Hamilton-Jacobi function. We are not allowed to set $k=-1$, but we can set $k$ arbitrarily close to $-1$ and so our phase $q\chi$ can be arbitrarily close to the Hamilton-Jacobi function. Let us keep $k$ arbitrary and consider a $k$ dependent Hamilton-Jacobi function
$$
S^{(k)}_{HJ} := q\chi.
$$
For $k=-1$ we obtain the usual Hamilton-Jacobi function, which we will denote by $S_{HJ}:=S_{HJ}^{(-1)}$. The relation between $S^{(k)}_{HJ}$ and $S_{HJ}$ is simply
\begin{equation} \label{Sk}
S^{(k)}_{HJ} = S_{HJ} + \left(1+\frac{1}{k}\right) m \int_{\gamma} \sqrt{u^2}d\tau.
\end{equation}
We shall use this relation later on. Now, in view of (\ref{chi}) the derivative of $S^{(k)}_{HJ}$ and the $k$ dependent Hamilton-Jacobi equation are
\begin{equation}  \label{hjeq}
\partial_{\mu}S^{(k)}_{HJ} = \frac{1}{k}p_{\mu} - qA'_{\mu}, \ \ \ \ (\partial S^{(k)}_{HJ} +qA')^2 = m^2/k^2
\end{equation}
so we are able to recover the usual Hamilton-Jacobi equation for $k=1$ or $k=-1$. Returning to the gauge transformation (\ref{gtr}),  we see that (\ref{hjeq}) results in
\begin{equation} \label{asquared}
q^2 A_{\mu}A^{\mu} = m^2/k^2.
\end{equation}
Consequences of this equation will be discussed in Section 6. Here let us only note that in view of (\ref{asquared}), the determinant of the metric (\ref{metric}) is
$$
\det g_{\mu\nu} = -1 -k\frac{q^2}{m^2}A^2 = -1- \frac{1}{k}
$$
and so $g_{\mu\nu}$ would change the signature if we had $k \in (-1,0)$. Now it is clear that the requirement $ k \in (-\infty,-1)\cup (0,\infty)$, which we derived earlier, is in fact equivalent to saying that the signature does not change.

Using the formula for $p_{\mu}$ in (\ref{hjeq}) and (\ref{gtr}) we also obtain
\begin{equation} \label{moment}
p_{\mu} = kq A_{\mu}
\end{equation}
which has to be interpreted as follows: the particle's direction is indicated by the field $A_{\mu}$. At this point it is appropriate to state that we have arrived at the equation (\ref{moment}) by using the equivalence principle for charged bodies and hence this equation must be regarded as the necessary condition. However, the LHS of (\ref{moment}) is an observable, while the RHS is proportional to the electromagnetic potential. Therefore we must conclude that $A_{\mu}$ is as physical as the momentum.

There are several important consequences of the equation (\ref{moment}).

\begin{itemize}
\item First, returning to (\ref{Sk}), we obtain another formula for the derivative of $S^{(k)}_{HJ}$; therefore, we have
\begin{equation} \label{dSk}
\partial_{\mu}S^{(k)}_{HJ} = \partial_{\mu}S_{HJ} + \left(1+\frac{1}{k}\right) p_{\mu}.
\end{equation}
However $\partial_{\mu}S_{HJ}$ should satisfy the usual Hamilton-Jacobi equation $(\partial S_{HJ} +qA')^2 = m^2$. Indeed, using (\ref{dSk}) we have
\begin{equation}  \label{dSk1}
\partial_{\mu} S_{HJ} +qA'_{\mu} = \partial_{\mu} S^{(k)}_{HJ} +qA'_{\mu} - \left(1+\frac{1}{k}\right) p_{\mu}  = -p_{\mu}
\end{equation}
where in the last step we used the definition of $A'_{\mu}$ via gauge transformation (\ref{gtr}) and substituted (\ref{moment}). Now, taking the square of (\ref{dSk1}), we arrive at the Hamilton-Jacobi equation for $S_{HJ}$.

\item Second, the equation (\ref{dSk1}) together with (\ref{moment}) imply that if $A'_{\mu}$ is not pure gauge, then the value of $k$ is unique. To see this, assume that there are two possible values, $k$ and $\bar{k}$. Substituting (\ref{moment}) into (\ref{dSk1}) and expressing $A_{\mu}$ in terms of $A'_{\mu}$ via (\ref{gtr}), we find that
\begin{equation} \label{dSk2}
\partial_{\mu} S_{HJ} +qA'_{\mu} = -kqA'_{\mu} -k\partial_{\mu}S^{(k)} = -\bar{k}qA'_{\mu} -\bar{k}\partial_{\mu}S^{(\bar{k})}
\end{equation}
where in the last step we used the fact that  $S_{HJ}$ and $A'_{\mu}$ are independent of $k$ and so the whole LHS of (\ref{dSk2}) is $k$ independent. This however implies that
$$
A'_{\mu} =  \partial_{\mu} \left(  \frac{kS^{(k)}-\bar{k}S^{(\bar{k})}}{q(\bar{k}-k)} \right).
$$
We have therefore arrived at the statement that potential $A'_{\mu}$ is pure gauge. This implies that our assumption about the existence of $k$ and $\bar{k}$ is incorrect hence $k$ is unique.

\item Third, introducing the generalised momentum $\pi^{\mu} = p^{\mu} + q A^{\mu}$ we see that
$$
\pi^{\mu} = p^{\mu}\left(1+\frac{1}{k}\right) \ \ \ \Longrightarrow \pi^2 =  \left(1+\frac{1}{k} \right)^2m^2
$$
so for $k=-1$ we would  have $\pi^{\mu}=0$.

\item Forth, by substituting (\ref{moment}) into the Lorentz-force law (\ref{eom}) we obtain a consistency condition for $A_{\mu}$, we have
$$
qk \dot{A}_{\mu} = qF_{\mu\nu}u^{\nu} \ \ \ \Longrightarrow (1+k)\dot{A}_{\mu} = \partial_{\mu}A_{\nu} u^{\nu}.
$$
We see again that $k=-1$ plays a special role. Contracting the above equations with $u^{\mu}$ we see that for an arbitrary $k$ we have
$$
u^{\mu}\dot{A}_{\mu} = u^{\mu}u^{\nu}\partial_{\nu}A_{\mu} = 0.
$$
However, for $k=-1$ the condition is even stronger
\begin{equation}  \label{Ap}
\partial_{\mu}A_{\nu} p^{\nu}=0 \ \ \ \ \Longrightarrow \ \ \ \ A^{\nu}\partial_{\mu}A_{\nu}=0,
\end{equation}
where we used  (\ref{moment}) in the last step.
\end{itemize}

\section{A normalisation choice}
So far we have performed the calculations not imposing normalisation constraints on $u^{\mu}$. In curved space-time we can always set $gu^2 = C_1$ while in Minkowski space we may set $u^2= C_2$, where $C_1$ and $C_2$ are constants. In our problem we should be able to set these conditions simultaneously since $u^{\mu}$ is the same 4-velocity from both curved space and from Minkowski space view.
That this is possible follows from the equation (\ref{proportional}) which implies that $C_1$ and $C_2$  are not independent, but satisfy $C_1 = C_2/\sqrt{1-kC^2)}$. In fact, we can turn this argument around and say that: the possibility that $C_1$ and $C_2$ can be set constant simultaneously \emph{implies}  that $\kappa/\sqrt{gu^2}=const.= C_3$. Therefore, the condition (\ref{condition}) may have already been deduced at the level of the action (\ref{action}) while consistency with the Lorentz force merely implies that $C_3=C$ (cp. (\ref{condition})).

It is clearly most convenient to set $u^2=1$ and we shall use this convention in the next section. 

\section{Consistency check}
The fact that condition (\ref{cons1}) involves both $u^{\mu}$ and $A_{\mu}$ is not a surprise since the geodesic equation is quadratic in four-velocities, while the Lorentz-force equation is linear in $u^{\mu}$. Here we investigate this directly by looking at the geodesic equation
\begin{equation} \label{geodesic}
\dot{u}^{\mu}  + \Gamma^{\mu}_{\alpha\beta} u^{\alpha} u^{\beta} =0.
\end{equation}
Moreover, considering  the remarks from the previous section, we will assume that $gu^2=const$. 

A priori it is not obvious that (\ref{geodesic}) is already equivalent to (\ref{kappa2}) since the Christoffel symbols in (\ref{geodesic}) involve the inverse metric $g^{\mu\nu}$, which was never used in the previous derivation. The inverse of (\ref{metric}) is
\begin{equation}  \label{inverse}
g^{\mu\nu} = \eta^{\mu\nu} - \frac{a^{\mu} a^{\nu}}{k+ a^2},  \ \ \ \ a^{\mu} := k \frac{q}{m}A^{\mu}
\end{equation}
so the Christoffel symbols are
\begin{equation} \label{chris}
\Gamma^{\mu}_{\alpha\beta}= \frac{1}{2k} \left( g^{\mu\nu} a_{(\alpha} f_{\beta)\nu} + g^{\mu\nu}a_{\nu}s_{\alpha\beta} \right),
\end{equation}
$$
f_{\mu\nu} := \partial_{[\mu}a_{\nu]}, \ \ \ \  s_{\mu\nu} := \partial_{(\mu}a_{\nu)}.
$$
Note that we could at this point take advantage of the condition $a^2=1$; however, this turns out not to be necessary (this is expected since we did not use this condition in Section 2 when deriving the Lorentz force law). Substituting (\ref{chris}) and (\ref{inverse}) into (\ref{geodesic}) we obtain
\begin{equation}  \label{check}
\dot{u}^{\mu} + \frac{1}{k}{f_{\beta}}^{\mu}u^{\beta}  - \frac{a^{\mu}}{k+a^2}\left( -\frac{1}{k} a^{\nu}f_{\beta\nu} u^{\beta} + \frac{1}{2}s_{\alpha\beta}u^{\alpha}u^{\beta} \right)=0,
\end{equation}
where we used $a\cdot u=1$. Now, we observe that
$$
a\cdot u=1 \ \ \ \ \Longrightarrow  \ \ \ \ \   \frac{1}{2}s_{\alpha\beta} u^{\alpha} u^{\beta} = -a^{\nu} \dot{u}_{\nu}
$$
hence, returning to the $A_{\mu}$ variables, we finally obtain
$$
\dot{u}^{\mu}  + \Gamma^{\mu}_{\alpha\beta} u^{\alpha} u^{\beta} = g^{\mu\nu}\left(\dot{u}_{\nu} - \frac{q}{m}F_{\nu\alpha}u^{\alpha}\right)=0.
$$

The above calculation simultaneously clarifies the following problem: We used the equivalence principle to argue that the Lorentz force should be derivable from the geodesic equation. However the inverse assertion should also be true, i.e. it should be possible to use the Lorentz force equations and some relation involving $g_{\mu\nu}$ and $A_{\mu}$ so that the resulting equation looks like the geodesic one. Clearly, such a relation should be as in (\ref{metric}). Then we can use the above calculation in reverse to arrive at the geodesic equation. (The only caveat in this argument is related to the question of whether every metric can be represented as in (\ref{metric}). Globally, this assertion is not true; however, locally - which is enough here - it is fairly justified since locally one can make an even stronger choice by introducing the Fermi normal co-ordinates. Note however that we need to maintain $k \notin [-1,0]$.)

\section{Einstein's equations}

Because the observer may describe his trajectory using either Lorentz force law with the field $A_{\mu}$ or the geodesic equation with the metric (\ref{metric}), it is interesting to see if some equivalence/connection between the (sourceless) Maxwell action and the (sourceless) Einstein-Hilbert action (with metric (\ref{metric})) can be made. Let us therefore investigate what are the vacuum Einstein equations if the metric (\ref{metric}) is used with a constraint $k^2q^2A^2=m^2$.

To simplify calculations, we will be working with the dimensionless field $a_{\mu}$ as in (\ref{inverse}), so that the metric and its inverse are
\begin{equation} \label{gig}
g_{\mu\nu}= \eta_{\mu\nu}+ \frac{1}{k}a_{\mu}a_{\nu}, \ \ \ \ g^{\mu\nu}= \eta^{\mu\nu}-\frac{1}{k+1} a^{\mu}a^{\nu},
\end{equation}
where we used $a^2=1$. The convention whereby the indices are raised/lowered by $\eta^{\mu\nu}/\eta_{\mu\nu}$ will be very useful in this section. 

The condition $a^2=1$ results in several identities. We have
\begin{equation} \label{id1}
a^{\mu}\partial_{\nu}a_{\mu} = 0,
\end{equation}
\begin{equation}  \label{id2}
-a_{\mu}f^{\mu\nu}f_{\nu\rho}a^{\rho} = a_{\mu}f^{\mu\nu}s_{\nu\rho}a^{\rho}=a_{\mu}s^{\mu\nu}s_{\nu\rho}a^{\rho} = (a\partial a)^2,
\end{equation}
where we use a shorthand notation $(a\partial a)^2 = (a^{\mu}\partial_{\mu} a^{\rho}) (a^{\nu}\partial_{\nu} a_{\rho}$).
Let us now express the Christoffel symbols in a convenient way as
\begin{equation} \label{christ}
\Gamma^{\mu}_{\nu\rho} = \frac{1}{2k}a_{(\nu}{f_{\rho)}}^{\mu} + \frac{1}{2(k+1)}a^{\mu}s_{\nu\rho} - \frac{1}{2k(k+1)}a^{\mu}a^{\sigma}a_{(\nu}f_{\rho)\sigma},
\end{equation}
where $f_{\mu\nu}$ and $s_{\mu\nu}$ are as in (\ref{chris}). Using (\ref{id1}) we observe that $\Gamma^{\mu}_{\mu\nu}=0$ and so only two terms contribute to the Ricci scalar, they are
$$
R =  g^{\mu\nu}\partial_{\rho} \Gamma^{\rho}_{\mu\nu}  - g^{\mu\nu} \Gamma^{\rho}_{\mu\sigma}\Gamma^{\sigma}_{\nu\rho}.
$$
Calculating these terms is straightforward using (\ref{gig}), (\ref{christ}) and identities (\ref{id1}), (\ref{id2}), although it is a bit lengthy. We find that
$$
 \eta^{\mu\nu} \Gamma^{\rho}_{\mu\sigma}\Gamma^{\sigma}_{\nu\rho} =  a^{\mu}a^{\nu} \Gamma^{\rho}_{\mu\sigma}\Gamma^{\sigma}_{\nu\rho} =  -\frac{1}{4k^2}f_{\mu\nu}f^{\mu\nu} + \frac{1}{2k^2}(a\partial a)^2,
 $$
$$
a^{\mu}a^{\nu}\partial_{\rho} \Gamma^{\rho}_{\mu\nu}  = \frac{1}{2k}f_{\mu\nu}f^{\mu\nu}- \frac{1}{k}(a\partial a)^2 + \frac{1}{k}\partial_{\mu}(a^{\nu}{f_{\nu}}^{\mu})
$$
and so, up to the irrelevant total derivative (note that $\sqrt{-g}=\sqrt{1+1/k}$), the curvature is
\begin{equation} \label{R}
R = -\frac{1}{4k(k+1)}f_{\mu\nu}f^{\mu\nu}+\frac{1}{2k(k+1)}(a\partial a)^2.
\end{equation}
Returning now to the $A_{\mu}$ variables, we should augment the final Lagrangian by a quadratic term in $A_{\mu}$ so that the constraint $k^2q^2A^2=m^2$ is incorporated at the action level. Finally we arrive at the following action principle
$$
S = \frac{1}{16\pi G} \int d^4x  \sqrt{-\det g} R \ \ \ \to
$$

\begin{equation} \hspace{-0.15cm}  \label{actionpr}
\sqrt{\frac{k}{k+1}}\frac{q^2}{16\pi G m^2} \int d^4x \left[ -\frac{1}{4}F_{\mu\nu}F^{\mu\nu}  + \frac{q^2k^2}{2m^2}(A\partial A)^2  + \lambda m^2 \left(k^2q^2A^2-m^2\right) \right]
\end{equation}

where $\lambda$ is a dimensionless Lagrange multiplier.  

The above action describes the gravitational field as seen locally by the charged particle with mass $m$, i.e. when the $g_{\mu\nu}$ field is replaced by $A_{\mu}$ using the map (\ref{metric}). However, since the Einstein-Hilbert action is non-renormalizable, the action (\ref{actionpr}) should also have the same issues. We observe that non-renormalizability of the action (\ref{actionpr}) is guaranteed by the $(A\partial A)^2 = -AFFA$ term. The existence of this term is therefore expected (at this point we can also turn this argument around and say that gravity cannot be renormalizable because of the $(A\partial A)^2$ term in (\ref{R})).

Moreover, this term is very small compared to the $F^2$ term and becomes significant when $A_{\mu} \sim m/qk$, i.e. when the field is strong (e.g. if we consider the Coulomb potential then the $(A\partial A)^2 $ term becomes non-negligible at distances of order $kq^2/m$ i.e. Compton wavelength times $kq^2/2\pi$). Therefore for weak fields we conclude that the Einstein-Hilbert action with metric (\ref{metric}) results in (approximately) Maxwell electrodynamics in Dirac gauge $A_{\mu}A^{\mu} = const.$

\section{Radiation reaction force}

In this section we would like to go further and apply our reasoning in the case of a radiating body. The thought experiment we consider should in principle apply also to cases in which  the charged elevator is accelerating and radiating with a rate which is not negligible. We will therefore search for a generalization of the metric (\ref{metric}) in such a way that the radiation reaction force appears naturally from the variation principle. 

\subsection{Landau-Lifshitz proposal }

It is very well known that the Lorentz force equation is only approximate due to the radiation emitted by the accelerated particles. If the radiation is small, then the Lorentz force equation may be used however in general a certain modification of the equation is needed to account for such an interaction between a particle and a field. A standard result in this regard is the Abraham-Lorentz-Dirac  force \cite{ALD}
\begin{equation} \label{ALD}
\dot{u}^{\mu} = \frac{q}{m}{F^{\mu}}_{\nu} u^{\nu} + a^{\mu}, \ \ \ \ a^{\mu} := \frac{2}{3}\frac{e^2}{m} (\ddot{u}^{\mu} + \dot{u}^2 u^{\mu})
\end{equation},
where ${F^{\mu}}_{\nu} = \eta_{\nu\alpha}F^{\mu\alpha} $. 

To remove the derivatives of the four-velocity $u^{\mu}$ in $a^{\mu}$ (which cause the problematic runaway solutions) Landau and Lifshitz proposed replacing every $\dot{u}^{\mu}$ in $a^{\mu}$ with $\frac{q}{m}{F^{\mu}}_{\nu} u^{\nu} $ which results in  \cite{LL}

\begin{equation} \label{LL}
a^{\mu} = \frac{2q^3}{3m^2}\partial_{\nu} {F^{\mu}}_{\alpha} u^{\nu}u^{\alpha} + \frac{2q^4}{3m^3} F^{\mu \nu}F_{\nu \alpha} u^{\alpha} +\frac{2q^4}{3m^3}(F_{\alpha\beta}u^{\beta})(F^{\alpha\sigma} u_{\sigma})u^{\mu}.
\end{equation}

 This choice of $a^{\mu}$ has many virtues, in particular it does not lead to runaway solutions \cite{norunaway} and can be derived using perturbation theory \cite{Wald}. It is also equivalent to (\ref{ALD})  up to leading terms \cite{approx}. For a fairly recent experimental status of the Landau-Lifshitz proposal see \cite{Rafelski}.

\subsection{Leading order terms}

The Landau-Lifshitz proposal (\ref{LL}) contains terms of order $1/m^2$ and $1/m^3$. One way to  recover the $1/m^2$ term is to augment the metric (\ref{metric})  by a term of the same order. The only possibility, on dimensional grounds, is to consider the following extension of (\ref{metric})

$$
g_{\mu\nu} = \eta_{\mu\nu} + S_{\mu\nu}, 
$$
{\small
\begin{equation} \label{metric2}
 S_{\mu\nu}  =  k\frac{q^2}{m^2}A_{\mu}A_{\nu} 
 + k_1 \frac{q^3}{2m^2} \partial_{(\mu}A_{\nu)} + k_2\frac{q^3}{m^2}\eta_{\mu\nu} \partial A + k_3 \frac{q^2}{m^2}\eta_{\mu\nu} A^2, 
\end{equation}
}
\noindent where $k$, $k_1$, $k_2$ and $k_3$ are  dimensionless  (note that in cgs units with $c=1$ we have $[q^2]=\hbox{gram}\cdot \hbox{cm}$ and $[qA_{\mu}] = \hbox{gram}$ hence the $q^3/m^2$ factor in the middle terms) and where $\partial A = \eta^{\mu\nu}\partial_{\mu}A_{\nu}$ and $A^2 = \eta^{\mu\nu}A_{\mu}A_{\nu}$.  Before analysing this special case let us first write equations of motion for the  general, symmetric tensor $S_{\mu\nu}$. The variation of the action gives the following equations of motion
\begin{equation}\label{eom}
(\eta_{\mu\alpha} + S_{\mu\alpha})\dot{u}^{\mu} = \left( \frac{1}{2}\partial_{\alpha}S_{\mu\nu} -  \partial_{\nu}S_{\mu\alpha} \right)u^{\mu}u^{\nu},
\end{equation}
where we set $g_{\mu\nu}u^{\mu}u^{\nu}=const$. These equations are of course geodesic equations written is a non-standard way.   Now, just as in Section 2, we can argue that on one hand it is possible to set $g_{\mu\nu}u^{\mu}u^{\nu}=const.$, as we just did, but on the other hand the resulting equations should be equivalent to those in Minkowski space where we can set $\eta_{\mu\nu}u^{\mu}u^{\nu}=1$.
Therefore, the following consistency constraint should be set
\begin{equation}  \label{cons2}
S_{\mu\nu}u^{\mu}u^{\nu} = const.=:D.
\end{equation}
This constraint is a proper generalization of (\ref{cons1}).  Let us also observe that since we can choose $\eta_{\mu\nu}u^{\mu}u^{\nu}=1$ we should also have $\eta_{\mu\nu}u^{\mu}\dot{u}^{\nu}=0$. This imposes no further conditions since after contracting (\ref{eom}) with $u^{\alpha}$  we find that $0=\eta_{\mu\nu}u^{\mu}\dot{u}^{\nu}=-\dot{D}/2$ i.e. condition (\ref{cons2}). 

Let us now return to the metric (\ref{metric2}). The corresponding equations of motion are now 
$$
\left(\eta_{\mu\alpha} + B\eta_{\mu\alpha} + \frac{k_1 q^3}{2m^2}\partial_{(\mu}A_{\alpha)} \right)\dot{u}^{\mu} =
$$
\begin{equation} \label{eoms}
 \frac{q}{m}k\kappa F_{\alpha\nu}u^{\nu} - \frac{kq}{m} \dot{\kappa}A_{\alpha} 
 - \frac{k_1 q^3}{2m^2} \partial_{\mu} \partial_{\nu}A_{\alpha} u^{\mu} u^{\nu} +\frac{1}{2} \partial_{\alpha} B - \dot{B}u_{\alpha}
\end{equation}
and the constraint (\ref{cons2}) can now be written as
$$
k\kappa^2+k_1 \rho + B = D,
$$

\begin{equation} \label{cons2s}
\kappa:= \frac{q}{m}A_{\mu}u^{\mu}, \ \ \ \ \rho :=  \frac{q^3}{2m^2} \partial_{(\mu}A_{\nu)}u^{\mu} u^{\nu},
\ \ \ \  B:= \frac{1}{m^2}( k_2 q^3\partial A + k_3 q^2 A^2), \end{equation}

\noindent where $D$ is a dimensionless constant. 
 Let us now split the term  $ \partial_{\nu} \partial_{\mu}A_{\alpha}$ in  (\ref{eoms})  into gauge-invariant and gauge-dependent parts
$$
\partial_{\mu}\partial_{\nu}A_{\alpha} =\frac{1}{2} \partial_{\mu}F_{\nu\alpha} + \frac{1}{2}\partial_{\mu}\partial_{(\nu}A_{\alpha)}
$$
so the equations of motion split accordingly
$$
\dot{u}_{\alpha} = \frac{q}{m}k\kappa F_{\alpha\nu}u^{\nu} - \frac{k_1 q^3}{4m^2} \partial_{\mu} F_{\nu\alpha} u^{\mu} u^{\nu} + h_{\alpha},
$$
$$
h_{\alpha} :=  - \frac{kq}{m} \dot{\kappa}A_{\alpha} -\frac{k_1 q^3}{2m^2}\partial_{(\mu}A_{\alpha)} \dot{u}^{\mu}  - \frac{k_1 q^3}{4m^2} \partial_{\mu} \partial_{(\nu}A_{\alpha)} u^{\mu} u^{\nu} 
$$
\begin{equation}  \label{eom2}
+\frac{1}{2} \partial_{\alpha} B - \dot{B}u_{\alpha} -B\dot{u}_{\alpha},
\end{equation}
where $h_{\alpha}$ is a collection of gauge dependent terms. Now, we observe that 
$$
h_{\alpha} u^{\alpha} = -\frac{1}{2}\frac{d}{d\tau }(k\kappa^2+k\rho+B)
$$
(which also follows from the remark below (\ref{cons2})) therefore the condition (\ref{cons2}) follows from demanding that $h_{\alpha}=0$, i.e. from requiring that (almost - see below) all gauge-dependent terms in the equation of motion disappear. Therefore instead of setting the constraint (\ref{cons2}) we will rather set $h_{\alpha}=0$. 

Equations of motion still contain the gauge-dependent factor $\kappa$ in front of the Lorentz force. However, up to terms of order $1/m^2$, this factor is constant since
\begin{equation}  \label{expans}
\kappa = \sqrt{D-k_1 \rho - B} = \sqrt{\frac{D}{k}} - \frac{1}{2\sqrt{Dk}}(k_1\rho+B) + \ldots \ \ .
\end{equation}
If we now substitute expansion (\ref{expans}) to equations of motion in (\ref{eom2}) and keep only the terms up to order $1/m^2$, we find that
\begin{equation}  \label{eom1}
\dot{u}_{\alpha} = \frac{q}{m}\sqrt{\frac{D}{k}} F_{\alpha\nu}u^{\nu}  - \frac{k_1 q^3}{4m^2} \partial_{\nu} F_{\mu\alpha} u^{\mu} u^{\nu} + h_{\alpha} + O(m^{-3}). 
\end{equation}
We see that in order to obtain the $1/m^2$ term of  the Landau-Lifshitz result (\ref{LL})  we need to set
$$
D/k=1, \ \ \ \ k_1 = -\frac{8}{3}, \ \ \ \ h_{\alpha}=0.
$$
Therefore, we are able to derive the leading term of the radiation reaction force from the equivalence principle.
Note that up to the dimensionless constant $k_1$, we have arrived at the correct $1/m^2$ expression using only gauge invariance ($h_{\alpha}=0$) and compatibility with the Lorentz force ($D/k=1$).

To conclude this section, we have shown that not only the Lorentz force, but also the leading terms of the radiation reaction force can be derived from the geodesic equation for a suitably chosen metric.  This metric turns out to be a fairly simple extension of (\ref{metric})
\begin{equation} \label{metric1}
g_{\mu\nu} = \eta_{\mu\nu} + k \frac{q^2}{m^2}A_{\mu}A_{\nu} - \frac{8}{6}\frac{q^3}{m^2}\partial_{(\mu}A_{\mu)}.
\end{equation}
The geodesic equation for the above metric results in the equation of motion
\begin{equation} \label{ALD1}
\dot{u}^{\mu} = \frac{q}{m}{F^{\mu}}_{\nu} u^{\nu} +\frac{2q^3}{3m^2}\partial_{\nu} {F^{\mu}}_{\alpha} u^{\nu}u^{\alpha}
\end{equation}
(as long as the condition $h_{\alpha}=0$ is met), i.e. the Lorentz force augmented by the leading term of the Landau-Lifshitz equation \cite{LL}.

\subsection{Higher order terms}

Clearly, one would like to find the metric for which all the terms of the Landau-Lifshitz equation appear. However, the third term of the equation is third-order in four velocities. This implies that in order to recover these terms one needs to apply the non-standard $u^{\mu}$ or $\dot{u}^{\mu}$ dependent metric. 

Non-standard metrics are completely acceptable in view of the problem we are considering. The generalization of the Lorentz force law given by the Landau-Lifshitz equation (\ref{LL}) is an attempt to include the influence of the particle's motion on the field. For small accelerations, the corrections are negligible, and one may use the Lorentz-force; for large accelerations one has to take into account corrections from the interaction between the particle and the field. One may say that the particle affects the background field in a way that is no longer negligible. Translating this into the gravitational field picture, it simply implies that the metric we are looking for is affected by the observer's motion. If this is the case, then clearly the metric may depend on the velocity and acceleration of the observer. 

However, there is a caveat related to this reasoning: we are assuming that the equivalence principle can be used (is valid) in situations in which the background field is affected by the particle. This would imply that gravitational self-force effects can be incorporated into the geodesic equation by considering the metric $g_{\mu\nu}+ h_{\mu\nu}$, where $g_{\mu\nu}$ is the background field and $h_{\mu\nu}$ is an appropriate correction from the gravitational self-force. This assertion, up to leading order, follows from the MiSaTaQuWa equations of motion \cite{MiSaTaQuWa}, therefore we conclude that our approach to consider non-standard metrics is well-grounded; however, this is out of the scope of the current paper.

\section{Summary}

It is commonly argued that the Lorentz force equation cannot be considered as a geodesic one due to the fact that it depends on the mass of the particle. There is however a caveat in this argument; namely, we can make the metric also depend on the mass. This, in fact, is a necessity, considering that such a metric should depend on the dimensional field $A_{\mu}$.

 If one wishes to obtain the equivalence between the geodesic and Lorentz force trajectories, it seems logical to take advantage of some characteristic feature of the electromagnetic field. In our opinion, this feature is Gauss's law, the consequence of which is the screening of the field inside a charged conducting body. This allows one to consider a modification of Einstein's elevator thought experiment in which the elevator is charged. The observer inside the elevator cannot detect the electromagnetic field and so the equivalence between the geodesics and the Lorentz force trajectories follows.

 We have derived the consistency condition (\ref{cons1}) from the requirement that the geodesic equation for the metric (\ref{metric}) coincides with the Lorentz force equation in Minkowski space. Such a condition can be achieved by choosing a certain gauge on the trajectory. Therefore, we must conclude that the equivalence principle, exploited in this way, fixes the gauge of the electromagnetic potential (albeit marginally).  

Working in this gauge and using the $A_{\mu}$ dependent metric (\ref{metric}) in the Einstein-Hilbert action results in an action that for weak fields (in particular for regions far away from the sources) coincides with Maxwell electrodynamics. We have therefore arrived at a certain correspondence already at a classical level, i.e. for weak fields equations for the electromagnetic field follow from Einstein's equations. As expected, for strong fields (in particular for small scales), the correspondence is broken by a term that makes the theory non-renormalizable.

Extending our approach to radiating bodies is possible as we were able to recover the leading terms of the radiation reaction. Recovery of all the terms would require the use of non-standard metrics and is beyond the scope of this paper.

\section{Acknowledgments}

Early version of this manuscript was developed at KTH and NORDITA. Their support, especially that of J. Hoppe, is greatly appreciated.
I would like to thank R. Janik and P. O. Mazur for their comments. This work is supported in part by the NCN grant UMO-2016/21/B/ST2/01492.


\begin{thebibliography}{0}

\bibitem{KK}
T. Kaluza, Sitzungsber. Preuss. Akad. Wiss. Berlin. (Math. Phys.) 966-972 (1921); \
O. Klein, Zeitschrift f\"ur Physik A 37 (12): 895-906 (1926).

\bibitem{ALD} 
P. A. M. Dirac, PRSL A167, 148-69 (1938), and references therein.

\bibitem{Rafelski} 
Y. Hadad, L. Labun, J. Rafelski, N. Elkina, C. Klier, and H. Ruhl, Phys. Rev. D 82, 096012.

\bibitem{LL} 
L. D. Landau and E. M. Lifshitz,  \textit{Teoria Pola}, PWN, Warszawa (2009), 4th edition.

\bibitem{norunaway} 
H. Spohn, Europhys. Lett. 49, 287-292 (2000).

\bibitem{Wald} 
S. E. Gralla, A. I. Harte, R. M. Wald, Phys. Rev. D 80, 024031 (2009).

\bibitem{approx} 
F. Rohrlich, Phys. Lett. A 303, 307 (2002).

\bibitem{MiSaTaQuWa}
Y. Mino, M. Sasaki, T. Tanaka, Phys. Rev. D, 55, 3457-3476, (1997); \ 
T. C. Quinn, R. M. Wald, Phys. Rev. D, 56, 3381-3394, (1997); \
S. E. Gralla, R. M. Wald, Class. Quant. Grav. 25:205009, 2008; Erratum-ibid.28: 159501, 2011; \ 
E. Poisson, A. Pound, I. Vega, Living Rev. Relativity 14,  (2011),  7.

\end{thebibliography}
\end{document}